# Cooperative Communication based Connectivity Recovery for UAV Networks

Wen Tian[1,2], Zhenzhen Jiao[1], Min Liu[1], Meng Zhang[3], Dong Li[3]

[1] State Key Laboratory of Computer Architecture, Institute of Computing Technology, Chinese Academy of Sciences, Beijing 100190, China
[2] University of Chinese Academy of Sciences, Beijing 100049, China
[3] RDA FOA ART-CN1, Corporate Technology, Siemens Ltd., China, Beijing 100102, China

*Abstract*—UAV networks often partition into separated clusters due to the high node and link dynamic. As a result, network connectivity recovery is an important issue in this area. Existing solutions always need excessive movement of nodes and thus lead to low recovery efficiency in terms of the time and energy consumption. In this paper, we for the first time study the issue of how to utilize cooperative communication technology to improve the connectivity recovery efficiency in UAV networks. We propose a Cooperative Communication based Connectivity Recovery algorithm for UAV Networks, named C$^3$RUN. The key novelty is C$^3$RUN not only uses cooperative communication to enlarge node's communication range and thus achieve quick repair of network connectivity, but also enables nodes to proactively move to better places for ensuring the establishment of cooperative communication links. We conduct extensive simulations to evaluate the performance of C$^3$RUN. The simulation results reveal that C$^3$RUN can not only achieve connectivity recovery with less nodes and shorter distance to move, but also always finish recovery with less time, when comparing with existing work. Furthermore, C$^3$RUN can achieve 100% success ratio for connectivity recovery.

## I. INTRODUCTION

Unmanned Aerial Vehicles (UAVs) have gained rapid development in recent years [1]. UAVs enable lots of new applications in both military and civilian areas because of their versatility, flexibility, easy installation and relatively small operating expenses. Such applications include search and destroy operations [1], border surveillance [2], wildfire monitoring [3], relay for ad-hoc networks [4,5], wind estimation [6], disaster monitoring [7], remote sensing [8] and traffic monitoring [9]. However, individual UAV often performs inefficiently when undertaking some complex tasks subject to its limited capability. Recently, the UAV swarm, which enables a team of UAVs to self-organize and work cooperatively to bring significant efficiency gain when undertaking tasks, has attracted increasingly attention. Cooperation between UAVs in a swarm can achieve higher task capacity than any individual capabilities of UAVs. However, the efficient usage of such self-organized UAV swarms still faces multiple challenges and one of the most prominent problems is how to guarantee efficient communication among the UAVs especially in dynamical and complex environments. Generally, communications among the UAVs in a self-organized swarm rely on a self-organized ad-hoc network of all UAVs. Unlike many other wireless networks, nodes in UAV networks may arrive and leave suddenly depending on applications or external influence, which may cause the links to be intermittently disrupted. This poses many challenges to the network. A key problem is the vanishing of nodes and links may lead to a significant change of network topology and even network partition. When network partition happens, the network needs to recover its connectivity as soon as possible. Since UAV networks have high partition frequency due to the node dynamic, efficient connectivity recovery capability is now an important requirement to the UAV networks.

In this paper, we study the issue of how to recover a topology's connectivity with the highest recovery efficiency when a UAV network partitions. Here, the recovery efficiency is measured by the sum of moving distances of all UAVs needed to move for topology repairing. The reason we use such metric is straightforward: shorter distance of UAVs need to move indicate a shorter recovery time and also less energy consumption. In this paper, for achieving the highest recovery efficiency, i.e., minimizing the nodes' moving distances, we introduce the cooperative communication technology in UAV network's connectivity recovery.

Cooperative communication (CC) utilizes neighbor nodes to work as helpers and simultaneously transmit independent copies of analogous data to a destination node so that the destination can combine partial signals of nodes and decode them. CC has emerged as a promising technique to address issues such as channel impairment, energy limitations and radio spectrum constraints [10-12]. Recently, research on topology control with CC in ad-hoc networks can been found in literatures which focused on network connectivity, path energy-efficiency, coverage extension and node transmission power reduction, e.g., [13][14]. However, existing work only considers CC in static scenarios. However, once a set of static nodes with CC cannot obtain enough communication range for reaching others, they will never get chance to connect to other again. In contrast, in this paper, we proposed a novel cooperative communication based connectivity recovery algorithm for UAV network, C$^3$RUN. In C$^3$RUN, nodes try to

This work was supported by the National Natural Science Foundation of China (No. 61732017, No. 61501125, No. 61472404, No. 61472402 and No. 61502457).

use CC to establish long-distance communication link between partitioned network parts to reduce movement of nodes. Once static CC failed, nodes can proactively move to better places for establishing CC links. To the best of our knowledge, this is the first work that utilized CC for connectivity recovery in UAV networks, especially nodes can proactively move to improve CC's efficiency.

The remainder of this paper is structured as follows. Section II introduces the background and related work. Section III presents the problem statement and algorithm design. Simulation results are provided in Section IV. In section V, we conclude this paper.

## II. BACKGROUND AND RELATED WORK

In this section, we first introduce the background related to UAV networks and cooperative communication. Next, we review the related work.

### A. Background

**UAV networks:** UAV networks have some similarities with Mobile Ad hoc Networks (MANETs) and Vehicular Ad hoc Networks (VANETs). However, unique characteristics also exist. For example, the topology dynamic of UAV network is much higher than others. Relative positions of UAVs may change not only in two but also in three dimensions with high speed. Network may partition because of node or link loss frequently. Thus, how to achieve efficient connectivity recovery is thus very important to UAV network and also its applications. Furthermore, energy constraint is another key issue in UAV network since UAV's energy not only affects its communication but also affects its flight time and task operating time.

**Cooperative communication:** cooperative communication has emerged as a promising technique to address issues such as channel impairment, energy limitations and radio spectrum constraints. In CC, by exploiting the broadcasting nature of wireless communication networks, neighbor nodes that overhear transmission between two nodes will enter a cooperation mode with source to process the overheard signal and forward it to the destination node. The signals received from the source and relay nodes can be combined to achieve higher diversity gain and more reliable transmission to overcome fading and channel impairments. Depending on how the relay node processes received signals, various signaling methods are proposed in literature, including Amplify and Forward (AF), Decode and Forward (DF) and Compress and Forward (CF). For more detail please refer to [15]. Since the SNR of signals at a receiver can be largely enhanced when helper's cooperative transmission exists, CC is often utilized for energy saving by reducing nodes' transmission power in ad-hoc networks in previous work. Cooperative Bridges in [16] utilizes CC to bridge disconnected network clusters by utilizing CC's transmission-distance expansion ability. Similarly, in this paper, we also utilize such ability of CC to improve connectivity recovery efficiency in UAV networks. In the following, we will review related work in the related fields.

### B. Related Work

Connectivity recovery from partitioned network is a key issue in ad-hoc networks and has been studied for years [17]. In this section, we will review some typical work which achieves connectivity recovery by utilizing node's mobility. DORMS proposed in [18] deployed relay nodes among disconnected network parts. By modeling the network topology as Steiner tree, DORMS can largely reduce the number of the required relay nodes. The work in [19] studied how to improve the fault-tolerance degree in ad-hoc robotic network. In [19], a block movement algorithm is proposed to eliminate cut-vertex-dependency in network. In this algorithm, leaf nodes will be scheduled to move toward cut-vertices to increase the connectivity degree around these cut-vertices and thus increases the whole network's degree of fault-tolerance. RIM in [20] is a distributed algorithm which achieves connectivity recovery by enabling nodes' to gather to a position. In RIM, when a node fails, its neighbors will move inward to its place so that they can connect with each other. The mechanism in RIM can also work in UAV networks. However, by RIM, nodes in network must perform continuous and cascaded motions. The relocation procedure is recursively applied to the related nodes until no one needs to move. LeDiR in [21] extended RIM by reducing the excessive movement in RIM. When cut-vertices failed and the network partitioned, LeDiR will choose the smallest node block to move and thus can minimize the number of the moved nodes. Although many algorithms have proposed for improving the connectivity recovery efficiency in wireless ad-hoc networks, research effect in UAV networks still lacks.

Cooperative communication has been utilized for connectivity maintenance and recovery in recent years. In [16], a novel topology control algorithm named Cooperative Bridges is proposed. In Cooperative Bridges, network partitions are first re-connected by enabling nodes to communicate with each other by using CC. To further reduce the energy consumption, a two-layer MST scheme is then used to remove the redundant links. In [22], the authors proposed an algorithm to establish k-fault-tolerant network by using CC to realize an energy-efficient fault-tolerant network. They established a *k*+1-connected network by selecting *k* help nodes for each node and use energy-efficient CC links to replace direct links. Previous work has revealed that CC can be very helpful for establishing a long transmission range and thus can achieve quick repair to network partition. However, existing work can be only used in static scenarios where node's position will never change. How to use CC in the high dynamic UAV networks, where partitioned network clusters may not be able to re-connect via directly usage of CC, is still an open issue.

## III. PROBLEM STATEMENT AND ALGORITHM DESIGN

In this section, we first introduce the system model. Next, we formulate the connectivity recovery problem under CC model. Finally, we propose the design details of C³RUN and also a brief discussion to its performance.

### A. System Model

In this paper, the UAV network under study can be modeled as a graph $G = (V, E)$, where $V(G)$ and $E(G)$ represent the sets of nodes and the set of links in the network, respectively. There are two kinds of links in our model: direct link and CC link.

**Direct link:** If the received average signal-to-noise ratio (SNR) at node $j$ from node $i$ is not less than a predefined threshold $\tau$, the receiver $j$ can successfully decode messages from the sender $i$, which means there exist a direct link from $i$ to $j$. To be specific, for a sender node $i$ to communicate with node $j$ directly, as shown in Fig. 1(a), the average SNR $\gamma_{ij}$ at receiving node $j$ from sending node $i$ should satisfy the following constraint:

$$\frac{P_i \mathbb{E}[|h_{ij}|^2]}{(d_{ij})^\alpha N} = \gamma_{ij} \geq \tau, \quad (1)$$

where $P_i$ denotes the transmission power of node $i$. $\tau$ denotes the SNR threshold required for correctly decoding a message. $\alpha$ is the path loss exponent. $h_{ij}$ is the channel coefficient from node $i$ to node $j$, which is generated by a Rayleigh distribution. $d_{ij}$ is the distance between node $i$ and $j$. $N$ is the noise power.

**CC link:** In cooperative communication, the neighbors of node $i$ can act as helper nodes to send the same packets simultaneously from node $i$ to node $j$. Node $j$ can then combine the received copies from sender node $i$ and its helper nodes to decode the original message. Here, use $\Omega_i$ to denote the set of a sender node $i$ and its helper nodes. If the total SNR received at node $j$ from $\Omega_i$ is not less than a threshold $\tau$, we can establish a CC link from node $i$ to node $j$, as shown by the following equation:

$$\sum_{i \in \Omega_i} \frac{P_i \mathbb{E}[|h_{ij}|^2]}{(d_{ij})^\alpha N} = \sum_{i \in \Omega_i} \gamma_{ij} \geq \tau, \quad (2)$$

According to (2), CC link can extend the transmission range of a node by utilizing the cooperative transmissions from its helpers, as shown in Fig. 1(b), which motivates us to utilize CC link to re-connect the partitioned network clusters with higher speed and lower moving distance than existing work. However, as we have introduced earlier, due to the high dynamic of UAV networks, only the existence of CC links cannot guarantee the connectivity recovery of the UAV network. How to utilize both CC and the nodal mobility to ensure the network connectivity to be recovered with high recovery efficiency is a nontrivial issue and will be formulated and discussed in the following sections.

### B. Problem Statement

The connectivity recovery problem under study can be formulated as follows.

Given a UAV network $G = (V, E)$ which has partitioned into multiple clusters because of failure of cut-vertex. For each pair of clusters, choose appropriate node $i$ and node $j$ in two clusters respectively to establish CC links between them. Let $\Omega_i$ denote the set of node $i$ and its helper nodes and $\Omega_j$ denote the set of node $j$ and its helper nodes, respectively. In cooperative communication, the total SNR at node $i$ from $\Omega_j$ is $\sum_{n \in \Omega_j} \frac{P_n}{(d_{ni})^\alpha}$, and the total SNR of node $j$ from $\Omega_i$ is $\sum_{m \in \Omega_i} \frac{P_m}{(d_{mj})^\alpha}$. To make clusters re-connected, some nodes may need to be relocated to satisfy the following formulas:

$$\sum_{n \in \Omega_j} \frac{P_n}{(d_{ni})^\alpha} \geq \tau, \quad (3)$$

$$\sum_{m \in \Omega_i} \frac{P_m}{(d_{mj})^\alpha} \geq \tau. \quad (4)$$

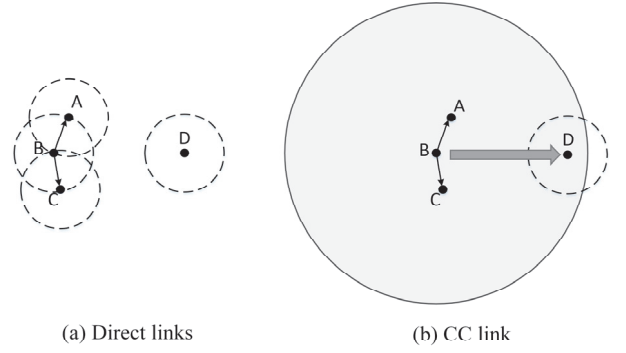

Fig. 1. Illustrations of direct link (a) and CC link (b).

Relocation of nodes in this procedure should never lead to new partitions in other positions in the network. Furthermore, to ensure such relocation of nodes achieves high efficiency, the total moving distance should be minimized. We formulate this problem as follows.

**Problem:** Given a UAV network $G = (V, E)$ which has partitioned into multiple clusters because of the failure of a cut-vertex. For each separated cluster, choose a set of nodes in the cluster to establish a CC link to other clusters. A certain number of nodes in this cluster needs to be moved in order to ensure that CC link can be established and the whole network can thus recover from splitting state. At this process, the total moving distance of all nodes, $\sum_{i \in n} S_i$, is minimized. Since this problem is NP-Hard, a heuristic algorithm is proposed in the following subsection.

### C. Algorithm Design

In this section, we propose C³RUN, a Cooperative Communication based Connectivity Recovery algorithm for UAV Networks. In C³RUN, when network partitions because of a cut-vertex's failure, nodes from different partitioned clusters will try to establish CC links towards the separated clusters. Once such tries failed, nodes and their helpers will then move towards the position where node failure happens. Such moving will perform iteratively until all moveable nodes have moved but still cannot enable the CC based communication to other clusters. At that time, the whole cluster may move to failure position. As an assumption, C³RUN needs nodes to exchange their positions periodically so that nodes can still know other nodes' positions even when they have partitioned. However, it is obvious that when separating time lapses, position information of nodes not in the same cluster will be out of date. To avoid such situation, the recovery process of $C^3$RUN will start immediately when network partition happens. Next, we will introduce $C^3$RUN in detail.

**Preliminary**: Each node exchanges and maintains its one and two-hop neighbors' information table, which contains the neighbor's ID and position. Network will partition to several disjoint clusters when a cut-vertex node fails.

**Step 1 (Failure detection):** The UAV nodes will periodically send heartbeat messages to their neighbors to announce their existence. Missing of several rounds heartbreak messages means a node failure happens. Next, the neighbors of the failing node will determine whether it is a cut-vertex by checking their neighbor information table. Once they found the failing node lead to loss of communication to other nodes, then the failing node is a cut-vertex and the network has been partitioned. The recovery process will then be performed.

**Step 2 (Use static CC link to repair):** To simplify the following introduction, we first consider the case that the network has partitioned into two clusters due to the node failure. We label the two separated clusters as cluster 1 and cluster 2, respectively. Further, the neighbors of the failure node at cluster 1 and 2 are denoted as $f_1$ and $f_2$, respectively. When the recovery process starts, the nodes in $f_1$ will try to establish CC links to the nodes in $f_2$ with the help of their own neighbors nodes (i.e., helper) and vice versa.

If we assume that there are $m$ nodes in $f_1$ and $n$ nodes in $f_2$, then there will exist $m \times n$ pairs of unidirectional CC links. If a pair of CC links both satisfy the formula (2), then a bidirectional CC link will be established. If there is at least one bidirectional CC link between two separated clusters, then nodes in two clusters can communicate again with each other and the goal for network connectivity recovery achieves. However, if there is no bidirectional CC link can be established between two clusters, then the recovery process will go to the next step.

**Step 3 (Use mobility enabled CC link to repair):** If the network has not recovered from node failure when Step 2 finished, nodes will be selected and move to make sure the whole network re-connect. To be specific, for each node $i$ in $f_1$, it will calculate the SNRs from $\Omega_i$ to each node in $f_2$, via equation (2). Node $i$ sums SNRs to nodes in $f_2$, and exchanges such value to other nodes in $f_1$. The node who achieves the largest sum of SNRs to nodes in the opposite cluster will then be chosen for establishing the CC link enabled by node mobility. The reason is the largest sum of SNRs indicates the least distance to the remote cluster. Such process will take place in different clusters independently. Here, we assume the node chosen from $f_1$ is $f_{1,i}$ and the node chosen from $f_2$ is $f_{2,j}$. From formula (2), it is obvious that the decrease of the distance between nodes will increase the total SNR. Thus, the node $f_{1,i}$ and $f_{2,j}$ will move toward the position of failed node, simultaneously. The two nodes will never stop until the two clusters are connected through CC links or the two nodes will move out of the communication range of their own neighbors. If a bidirectional CC link still cannot be established when the

| Algorithm 1 $C^3$RUN |
| --- |
| 1  IF UAV $J$ detects a failure of its neighbor $F$ |
| 2    IF $F$ is a cut-vertex UAV |
| 3      CCLinksEstablish($J$); |
| 4      IF Check_NetworkPartitioned($J$) |
| 5        ChooseOneNode($J$); |
| 6        IF $J$ is selected |
| 7          MoveToFailedNode($J$); |
| 8          IF Check_NetworkPartitioned($J$) |
| 9            HelperNodesMove ($J$); |
| 10           IF Check_NetworkPartitioned($J$) |
| 11             MoveToFailedNode($J$); |
| 12             OtherNodesMove($J$); |
| 13           END IF |
| 14         END IF |
| 15       END IF |
| 16     END IF |
| 17   END IF |
| 18 END IF |

**Check_NetworkPartitioned($J$) {**
19  FOR each UAV in the Neighbor Information Table
20    $J$ checks the communication to UAV;
21    IF the communication lost
22      RETURE TRUE;
23    END IF
24  END FOR
25  RETURN FALSE; }

**CCLinksEstablish($J$) {**
26 Establish CC links to other neighbors of $F$ with the helper nodes. }
**ChooseOneNode($J$) {**
27 Choose the node among neighbors of $F$ which achieves the largest SNR. }

two nodes have arrived at the edges of their neighbors' communication range, the next phase will start as follows.

The next phase is to move the helper nodes of $f_{1,i}$ and $f_{2,j}$. Their helper nodes will move towards the position of failure node one by one until a bidirectional CC link is established or they would move out of the communication range of $f_{1,i}$ or $f_{2,j}$. If $f_{1,i}$ and $f_{2,j}$ still can't establish a bidirectional CC link after all their helpers have moved, then Step 2 will be executed once more. After that, if the network connectivity still cannot recover, the algorithm will go to the last step.

**Step 4 (Move other nodes in cluster):** The main idea of this step is to move all the nodes in a cluster to move towards the failure position at last for guaranteeing network connectivity recovery. However, this process is performed in an iterative manner as follows. Use $B_1$ to denote the set of the candidate nodes to move in cluster 1. Initially, $B_1$ only contains $f_{1,i}$ and its neighbors, which means $B_1 = \{\Omega_{f_{1,i}}\}$. When step4 starts, the neighboring nodes of $B_1$ will be added into $B_1$ one by one. All the nodes in $B_1$ will move to the failure position when $B_1$ is not a cut set. Such movement will never stop until the bidirectional CC link is established or $B_1$ becomes a cut set due to the movement. At that time, more nodes will be added into $B_1$. This process will be repeated until all the nodes in cluster 1 have been added into $B_1$. Then we can move $B_1$ towards failure position until two clusters communicate with each other. Here we should note that we introduce the algorithm based on an example where network partitions into two clusters. However, $C^3RUN$ can also work when network partitions into more clusters since $C^3RUN$ can be executed in multiple rounds to achieve cluster-by-cluster based repairing.

More details of $C^3RUN$ can be found in Algorithm 1.

## IV. PERFORMANCE EVALUATION

In this section, we conduct extensive simulations to evaluate the performance of $C^3RUN$ by comparing it with existing algorithm RIM [20], LeDiR [21] and Cooperative Bridges [22]. In our simulation, multiple UAVs are placed in a 300m×300m area to form a connected UAV network. The number of UAVs varies from 20 to 50. The speed of each UAV is set to 1m/s. The transmission range of each UAV is set to 50 meters. In each simulation, the failure UAV is randomly chosen. The following metrics are used to evaluate the performance of our algorithm:

**(1) The number of UAVs needed to move for connectivity recovery:** smaller number indicates a lighter change to the network topology.

**(2) The sum of moving distances needed by UAVs in recovery process:** smaller moving distance of UAVs also represents a lighter change to the network topology. Furthermore, it also leads to less energy consumption of each UAV and thus prolongs the network lifetime.

**(3) The time spent in the recovery process:** the time length for recovery is one of the most direct metrics for measuring the recovery efficiency.

**(4) The success ratio of the algorithm for recovering in different topologies:** such ratio measures whether an algorithm can always recover the network connectivity in different network topologies.

We compare the performance of $C^3RUN$ with RIM, LeDiR and Cooperative Bridges. As we introduced earlier, RIM utilizes the mobility of nodes to recover the network. In RIM, once a node failure happens, all other nodes move to its position successively. In contrast, LeDiR only chooses the cluster with the least number of nodes to move and thus can reduce the number of nodes need to be moved. Cooperative Bridges is a topology control scheme used in static ad-hoc networks to connect partitioned clusters by using CC. For each network density (i.e., node number), 15 times simulations are performed with randomly generated topologies.

Fig. 2 compares the number of moving nodes needed for connectivity repairing. From Fig. 2 we can find that, RIM needs more nodes to move than others. LeDiR always chooses the smallest cluster to move so that it outperforms RIM in this aspect. Furthermore, it is apparently that $C^3RUN$ achieves the best performance since it needs fewer nodes to move than both RIM and LeDiR, which can even achieve the recovery without any nodal movement when network is dense.

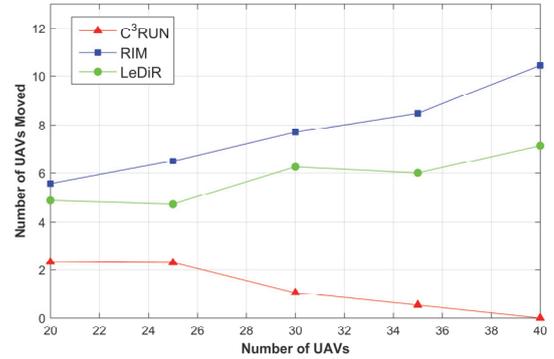

Fig. 2 Comparison of the number of UAVs needed to move.

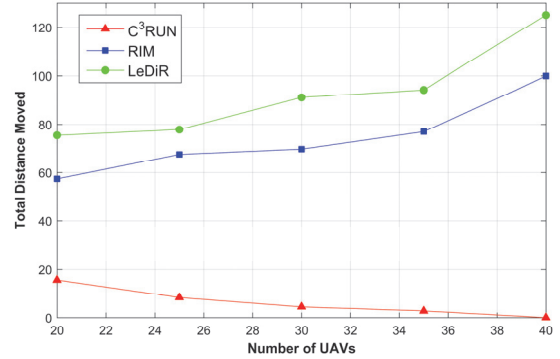

Fig. 3 Comparison of the total moving distance.

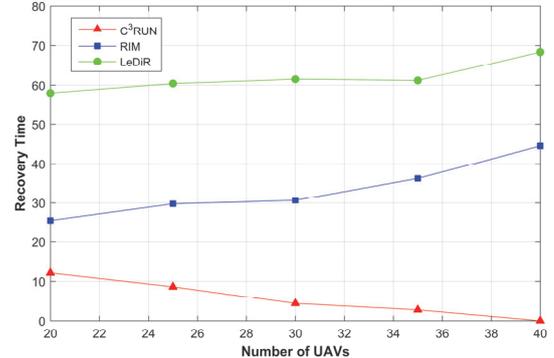

Fig. 4 Comparison of the time spent in recovery process.

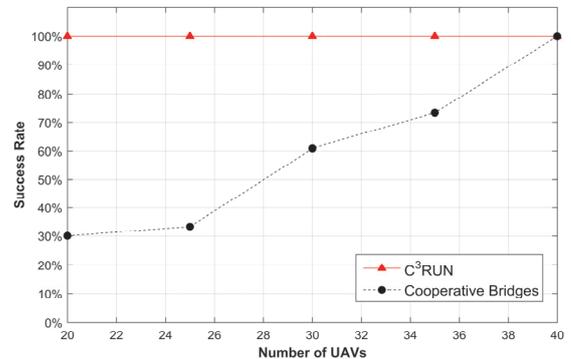

Fig. 5 Comparison of the success rate.

Fig. 3 shows C³RUN needs less movement of nodes for connectivity repairing than existing work. This can be attributed to two reasons, i.e., fewer nodes needed to move and larger nodal communication range brought by CC. Although LeDiR moves fewer nodes than RIM, it may lead small cluster but with long distance from failure position to move and thus increases the moving distance of each node. As a result, RIM outperforms LeDiR in this aspect.

In Fig. 4 we can find that C³RUN consumes less time to recover the network connectivity than existing work. In Fig. 3, when network density increases, the time length for recovery decreases. The reason is nodes can have more opportunities to obtain neighbors' help for establishing CC links when network density increases.

In Fig. 5, we compare C³RUN with Cooperative Bridges in terms of the success ratio for connectivity recovery. Fig. 5 reveals that static CC is also useful in some cases, i.e., when network density is high enough. However, in most cases, only CC is not enough for achieving a success recovery, which illustrates the necessity of C³RUN for enabling a mobility enabled CC in UAV networks.

## V. Concluding Remarks

In this paper, we proposed a cooperative communication based connectivity recovery algorithm for UAV networks, named C³RUN. C³RUN uses cooperative communication to enlarge node's communication range and thus achieves quick repair of network connectivity. In C³RUN, node's active movement is utilized to ensure that cooperative communication links can be always established when other methods fail. Extensive simulations were conducted to evaluate the performance of C³RUN. The simulation results revealed that, when comparing with existing work, C³RUN can not only achieve connectivity recovery with less nodes and shorter distance to move, but also always finish recovery with less time. Besides, C³RUN can achieve 100% success ratio for connectivity recovery.